# Impact of $Li_{2.9}B_{0.9}S_{0.1}O_{3.1}$ glass additive on the structure and electrical properties of the LATP-based ceramics

K. Kwatek[1*], W. Ślubowska[1], J. Trébosc[2,3], O. Lafon[2,4], J.L. Nowiński[1]

[1] Warsaw University of Technology, Faculty of Physics, 00-662 Warsaw, Poland

[2] Univ. Lille, CNRS, Centrale Lille, ENSCL, Univ. Artois, UMR 8181 - UCCS - Unité de Catalyse et Chimie du Solide, F-59000 Lille, France

[3] Univ. Lille, CNRS-FR2638, Fédération Chevreul, F-59000 Lille, France

[4] Institut Universitaire de France, 1 rue Descartes, F-75231 Paris Cedex 05, France

*Corresponding author. E-mail address: konrad.kwatek@pw.edu.pl

**Abstract**

The existing solid electrolytes for lithium ion batteries suffer from low total ionic conductivity, which restricts its usefulness for the lithium-ion battery technology. Among them, the NASICON-based materials, such as $Li_{1.3}Al_{0.3}Ti_{1.7}(PO_4)_3$ (LATP) exhibit low total ionic conductivity due to highly resistant grain boundary phase. One of the possible approaches to efficiently enhance their total ionic conductivity is the formation of a composite material. Herein, the $Li_{2.9}B_{0.9}S_{0.1}O_{3.1}$ glass, called LBSO hereafter, was chosen as an additive material to improve the ionic properties of the ceramic $Li_{1.3}Al_{0.3}Ti_{1.7}(PO_4)_3$ base material. The properties of this $Li_{1.3}Al_{0.3}Ti_{1.7}(PO_4)_3 - xLi_{2.9}B_{0.9}S_{0.1}O_{3.1}$ ($0 \leq x \leq 0.3$) system have been studied by means of high temperature X-ray diffractometry (HTXRD), [7]Li, [11]B, [27]Al and [31]P magic angle spinning nuclear magnetic resonance spectroscopy (MAS NMR), thermogravimetry (TG), scanning electron microscopy (SEM), impedance spectroscopy (IS) and density methods. We show here that the introduction of the foreign LBSO phase enhances their electric properties. This study reveals several interesting correlations between the apparent density, the microstructure, the composition, the sintering temperature and the ionic conductivity. Moreover, the electrical properties of the composites will be discussed in the terms of the brick-layer model (BLM). The highest value of $\sigma_{tot} = 1.5 \times 10^{-4}$ S·cm$^{-1}$ has been obtained for LATP–0.1LBSO material sintered at 800°C.

*1. Introduction*

The application of solid lithium ion electrolytes in the Li-ion battery (LIB) technology could offer many improvements related to safety, lifetime, specific power and energy. It is believed that all-solid-state Li-ion batteries (ASSLIB) may benefit from improved mechanical and thermal properties. Another advantage for the usage of solid electrolytes is their wider electrochemical window. It enables enhancing the power density of the batteries via application of new high-voltage (> 5 V) cathode materials [1-4]. The main criterion so that a material can be used as a solid electrolyte in the LIB technology is a high value of lithium ion conductivity exceeding $10^{-4}$ S·cm$^{-1}$ [1-5]. Unfortunately, most of the materials which fulfill this condition suffer from poor chemical stability against moisture. Among them, we can distinguish the glasses based on Li$_2$S [1-4]. On the other hand, the second group of materials, formed from the metal oxides exhibit high chemical stability, but their total ionic conductivity is relatively low [1-5]. Therefore, a major research direction is to find materials which are stable against moisture, while satisfying the conductivity criterion.

Within the oxide-based group of materials, one of the promising candidate is lithium titanium phosphate (LTP) with a chemical formula – $LiTi_2(PO_4)_3$ [6-13]. It crystallizes in NASICON-type structure with rhombohedral symmetry and belongs to R-3c space group. A number of its favorable properties, i.e. good thermal, electrochemical and mechanical stability as well as non-flammability, are of high interest for the ASSLIB technology. The bulk region of the LTP exhibits high lithium ion conductivity in the range of $10^{-4} - 10^{-3}$ S·cm$^{-1}$ at room temperature, depending on a processing method [6-13]. However, for the use of this material as solid electrolyte, not only the bulk conductivity but the total ionic conductivity should be taken into account. It has been already reported that the total ionic conductivity for polycrystalline or ceramic LTP is relatively low, ca. $10^{-8} - 10^{-6}$ S·cm$^{-1}$ at room temperature, due to a highly resistant grain boundary phase [6-13]. For this purpose, many efforts were made to improve the electric properties of the lithium titanium phosphate. One of the most efficient and commonly used strategy for enhancing the total conductivity of the $LiTi_2(PO_4)_3$ material consists in substituting a fraction of $Ti^{4+}$ ions by $Al^{3+}$ ones. The composition $Li_{1+x}Al_xTi_{2-x}(PO_4)_3$, where x = 0.3, called LATP hereafter, received much attention as it exhibits the best ionic properties among the NASICON-based family. Nevertheless, a discrepancy between the bulk and total conductivities has been observed and the total ionic conductivity notably depends on the preparation protocol. Apart from the conventional solid state reaction followed by sintering, LATP may be obtained via melt-quenching and sol-gel methods yielding the total conductivity as high as $10^{-3}$ S·cm$^{-1}$ at room temperature [14-20]. Even though both methods are efficient and reliable, they present certain disadvantages. First one requires high temperatures for melting the substrates, which exceed 1300°C, while the latter one focuses on synthesis based on highly expensive reagents, which prevents its industrial use.

On the other hand, as concerns the preparation of the ceramic material using conventional solid-state reaction method, the total ionic conductivity is significantly lower and ranges from $10^{-6}$ to $10^{-4}$ S·cm$^{-1}$ at room temperature [14, 21-23]. Therefore, to obtain highly conductive material, another approach has been proposed. It is based on the addition of some foreign phase into the base matrix in order to lower the grain boundary resistance [24-33].

The essential role of the foreign phase to enhance the total ionic conductivity of the parent base matrix has been studied for many different materials. Usually, the solid electrolyte materials are formed as ceramics, which require heat treatment at high temperatures. Hence, one of the best additive materials is supposed to be a glass with relatively low melting point. During sintering, glass should transform into a liquid phase, which can efficiently both cover the grains and fill the pores. The glass might also facilitate sintering process, and as a result, leads to lower processing temperature for the ceramic composite. One of the most popular glass used as an additive material seems to be $Li_3BO_3$, denoted later as LBO, due to its low glass transition and melting temperatures. N. C. Rosero-Navarro *et al.* [31,32] studied the composites based on garnet-type material and formed in the system $Li_7La_3Zr_xM_{1-x}O_{12}$ - $Li_3BO_3$ (M = Nb, Ta). They reported that the value of ionic conductivity of ceramic $Li_7La_3Zr_2O_{12}$ (LLZO) sintered at 900°C is relatively low, ca. $1 \times 10^{-7}$ S·cm$^{-1}$. On the other hand, when that material is sintered at 1200°C, it exhibits a significantly higher conductivity (ca. $1 \times 10^{-4}$ S·cm$^{-1}$). It has been found that LBO glass could play an essential role in lowering the sintering temperature (to 900°C), while maintaining good ionic properties. They explained that such good values of total ionic conductivity could be ascribed to the densification process, which takes place during the sintering of composite.

In our previous works [27,28], we investigated the impact of glassy $Li_3BO_3$ or $Li_{2.9}B_{0.9}S_{0.1}O_{3.1}$ (LBSO) additives into the LTP base matrix and we reported a significant improvement of the total ionic conductivity (by at least three orders of magnitude). Furthermore, N. Sharma *et al.* [30] also reported that LTP-based composite, to which $60Li_2SO_4$–$40LiPO_3$ glass is added, exhibits an increase in the total ionic conductivity by more than one order of magnitude in comparison to ceramic LTP. Authors propose that lithium ions present in the glassy phase may play an important role in the enhancement of grain boundary conductivity. In ref. [25], T. Hupfer presented the results of the composite formed in the system of LATP–LiTiOPO$_4$. They also reported that the introduction of foreign phase into the LATP base matrix might be an efficient way to lower the sintering temperature and to increase the density of the material, yielding improved electric properties of the material.

In this study, we present and discuss an alternative approach to form highly conductive composite materials based on LATP with the addition of Li$_{2.9}$B$_{0.9}$S$_{0.1}$O$_{3.1}$ glass. The use of LBSO as sintering aid was dictated by its relatively good ionic conductivity and low melting temperature. Similar concept was successfully applied in ref. [27] where LTP was used as a base matrix. The enhancement of total ionic conductivity by over three orders of magnitude was observed for LTP–LBSO composites. In this study we apply similar approach to the LATP–LBSO system in the view of improving its electrical properties. Our special interest is focused on the phase analysis by means of HTXRD and MAS NMR spectroscopy.

## 2. Experimental

The polycrystalline Li$_{1.3}$Al$_{0.3}$Ti$_{1.7}$(PO$_4$)$_3$ was obtained by means of solid-state reaction method. Reagent–grade chemicals, Li$_2$CO$_3$ (Sigma Aldrich), NH$_4$H$_2$PO$_4$ (POCh), anatase TiO$_2$ (Sigma Aldrich) and Al$_2$O$_3$ (Sigma Aldrich), were weighted in stoichiometric amounts, ground with a mortar and pestle and synthesized in an alumina crucible at 900°C for 10 h to yield the the final LATP compound. Li$_{2.9}$B$_{0.9}$S$_{0.1}$O$_{3.1}$ glass was obtained via the standard melt–quenching method. Stoichiometric amounts of reagents: Li$_2$CO$_3$ (Sigma Aldrich), H$_3$BO$_3$ (POCh) and Li$_2$SO$_4$ (Sigma Aldrich) were weighted, ground with a mortar and pestle, placed in an alumina crucible and annealed at 1100°C for 15 min. The molten mixture of reagents was then quenched between two stainless steel plates. Next, the obtained glassy plates were ball-milled with rotation speed of 400 rpm for 1 h in a planetary mill Fritsch Pulverisette 7. Subsequently, the obtained LBSO powder was added in stoichiometric amounts to as-prepared LATP material in molar ratio varying from 10 to 30%. Both powders were ball–milled in ethanol at 400 rpm for 1 h, dried and pelletized under uniaxial 10 MPa pressure. Finally, pellets of 6 mm in diameter and ca. 2 mm thick, were then formed and sintered at 700, 800 or 900°C for 2 h.

The phase composition of the as-synthesized materials and composite powders were examined by means of X-ray diffraction method. Data were collected in the 2θ range of 10° – 90° with 0.05° step and a count rate of 0.5 s at each step with CuKα line by means of Philips X'Pert Pro diffractometer. Additionally, temperature dependent XRD (HTXRD) was performed using Anton Paar HTK−1200 oven in the temperature 30 – 900°C range.

The thermal stability of the composites was determined by thermal gravimetric analysis (TGA) using TA Instruments Q600 calorimeter to register the mass loss and the temperature difference (with respect to the refence alumina crucible) during heating at the constant rate of 10 °C·min$^{-1}$ under air flow in the temperature range from 50 to 900°C. The measurements were performed at a heating rate of 10 °C.min$^{-1}$ on powdered samples of ca. 20 mg each.

The apparent density of the composites was determined using Archimedes method with isobutanol as an immersion liquid. The accuracy of this method was estimated as ca. 1%. The microstructure of freshly fractured pellets was investigated by means of scanning electron microscopy (SEM) employing Raith eLINE plus.

For impedance spectroscopy measurements, both bases of the as-formed pellets were polished and covered with graphite as electrodes. Impedance investigations were carried out employing Solatron 1260 frequency analyzer in a frequency range from 10$^{-1}$ to 10$^7$ Hz in the temperature range from 30 to 100°C, both during heating and cooling runs.

MAS NMR spectra were acquired on a 400 MHz Bruker Avance II NMR spectrometer equipped with a 4 mm HXY probe used in double resonance mode. The $^7$Li, $^{27}$Al and $^{31}$P NMR spectra were acquired at room temperature (RT) and a MAS frequency of 10 kHz using single-pulse experiments with pulse lengths of 1.0, 1.0 and 2.5 µs, radiofrequency (rf) nutation frequencies of 115, 60 and 50 kHz, respectively. The $^7$Li, $^{27}$Al and $^{31}$P NMR spectra result from the averaging of 512, 128, 8 transients using relaxation delays of 0.3, 1 or 20 s. The $^{11}$B NMR spectra were acquired on a 800 MHz Bruker Avance NEO NMR spectrometer equipped with a 3.2 mm XY probe free of boron background. The $^{11}$B NMR spectra were collected at room temperature and a MAS frequency of 20 kHz using single-pulse experiments with a pulse length of 1.55 µs and an rf nutation frequency of 40 kHz. The $^{11}$B NMR spectra resulted from averaging 1024 transients with a relaxation delay of 0.2 s. The $^7$Li, $^{27}$Al, $^{31}$P and $^{11}$B isotropic chemical shifts were referenced to 1 mol.L$^{-1}$ LiCl, 1 mol.L$^{-1}$ AlCl$_3$, 85 wt% H$_3$PO$_4$ aqueous solutions and neat (C$_2$H$_5$)$_2$O·BF$_3$ employing NaBH$_4$ solid ($\delta_{iso}$ = −42 ppm) as a secondary reference. The NMR spectra were simulated using dmfit software [34]. The $^{31}$P and $^{27}$Al NMR spectra were simulated assuming Gaussian lineshapes and considering only the isotropic shifts, whereas the $^7$Li NMR spectra were simulated assuming Lorentzian lineshapes for all bands and considering isotropic chemical shift and quadrupolar interaction. The $^{11}$B NMR signals of BO$_3$ and BO$_4$ sites were simulated as the sum of Gaussian and Lorentzian lineshapes.

## 3. Results and discussion

### 3.1 X–ray diffraction

The X-ray diffraction patterns of the as-prepared polycrystalline LATP and a composite containing 0.2 mol of Li$_{2.9}$B$_{0.9}$S$_{0.1}$O$_{3.1}$ glass (before and after sintering at 900°C for 2 h) are shown in Fig. 1. For LATP powder, the position and relative intensities of the main reflections agree well with the NASICON-like compounds with R-3c symmetry group (ICDD 00-035-0754). We observe some additional weak reflections at $2\theta$ angles equal to 20.6° and 26.3°, which are ascribed to TiO$_2$ anatase and AlPO$_4$ berlinite phases [8, 13–16, 19, 22, 24, 33, 35].

The non-sintered composite shows the same pattern profile as LATP. The XRD pattern for LATP–0.2LBSO composite sintered at 900°C consists mainly of the lines attributed to LATP. However after sintering of composite, additional reflections appeared at 17.0°, 18.5°, 20.4°, 22.2°, 27.0°, 27.3°, 27.7°, 31.0° and 39.3°. They were identified as $LiTiPO_5$ and $Li_4P_2O_7$ phases [8, 11, 24, 27–29, 33]. As evidenced by HTXRD studies, the most pronounced changes in phase composition could be observed above 700°C. At that temperature, peaks related to $TiO_2$ and $AlPO_4$ phases start to fade out while those attributed to the $LiTiPO_5$ and $Li_4P_2O_7$ appear. HTXRD results might suggest that at high temperatures the secondary phases $TiO_2$ and $AlPO_4$ phases are prone to react with either LATP or glass component, yielding the formation of new compounds $LiTiPO_5$ and $Li_4P_2O_7$.

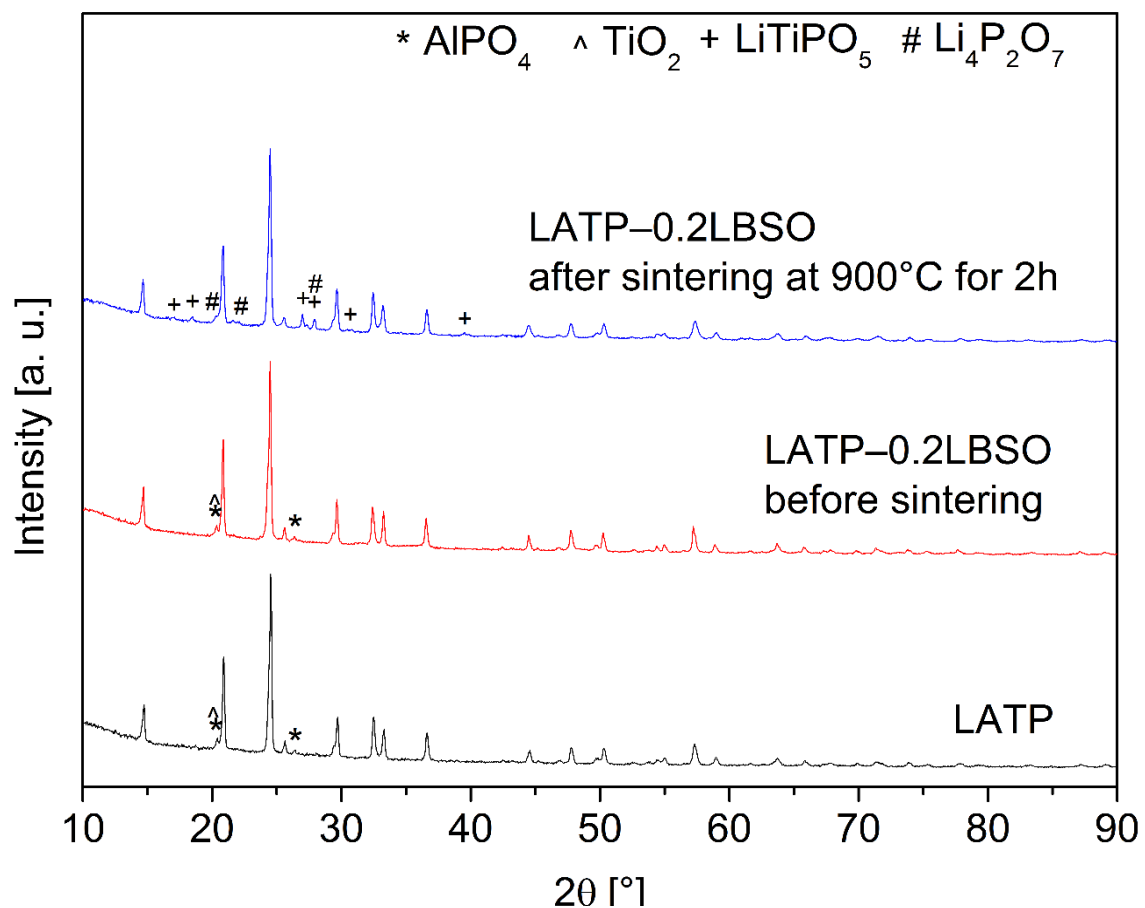

Figure 1 XRD patterns of the as-prepared LATP as well as the LATP–0.2LBSO composite before and after sintering at 900°C.

*3.2 Thermogravimetry*

Thermogravimetry investigations of the non-sintered composites, depicted in Fig. 2, reveal the loss of mass of ca. 3 wt.% for all LATP–LBSO compositions under study. Such a loss proceeds up to 350°C. Most probably, it is the result of the evaporation of moisture and residual ethanol adsorbed on grain boundaries. Above 350°C practically no further loss in mass is detected.

No thermal events are discernible on DTA thermograms as the LATP-LBSO composites are heated up to 900°C. It seems to be the consequence of the low glass content which is below the calorimeter's sensitivity threshold. As for the LBSO glass, it exhibits the glass transition at $T_g$ = 340°C, two crystallization peaks at $T_{c1}$ = 430°C and at $T_{c2}$ = 530°C and the melting point at $T_m$ = 735°C. More detailed discussion on the LBSO thermal properties can be found in [27].

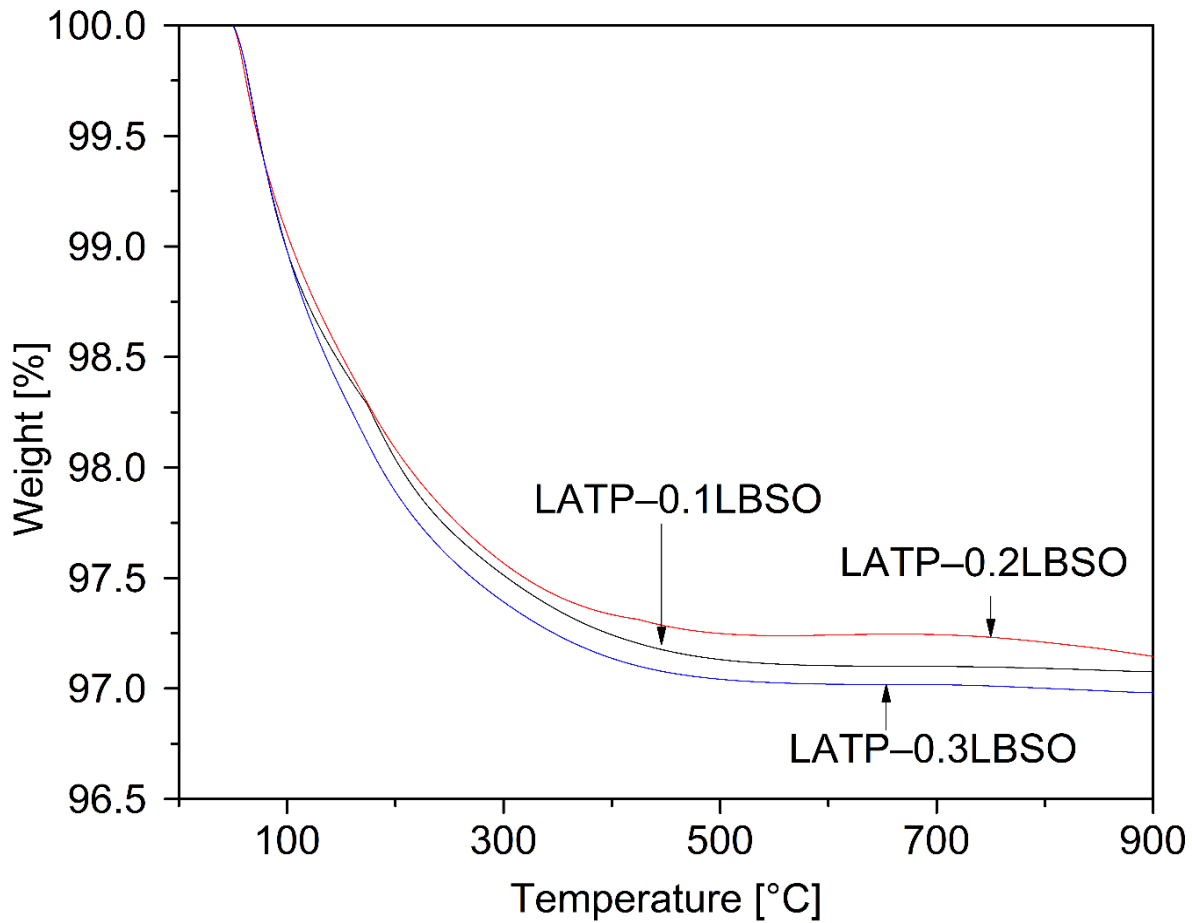

Figure 2 TG traces for LATP composites formed with various molar ratios of LBSO.

It is worth mentioning that the estimation of the melting point is crucial for this study, because it defines the lowest possible temperature at which the LATP-LBSO composites may be sintered. It is assumed that above that temperature, as the glass melts, the liquid phase efficiently fills in the pores of the LATP ceramics and facilitates its grain growth.

### *3.3 Microstructure and apparent density*

The SEM images of freshly fractured pellets for LATP sintered at 900°C (Fig. 3A) and composites with 0.1 mol of LBSO heat-treated at 700°C (Fig. 3B), 800°C (Fig. 3C) and 900°C (Fig. 3D) were analyzed in order to investigate the impact of LBSO additive on their microstructure. For sintered LATP, two grain fractions can be easily discerned, one of typical grain sizes of ca. 1 µm and other ones with a few micrometers in size. Some voids, microcracks and grain boundaries are also visible (Fig. 3A). LATP–0.1LBSO composite sintered at 700°C (Fig. 3B) exhibits similar grain size distribution but a higher porosity than LATP ceramic. As for a composite sintered at 800°C (Fig. 3C), the microstructure is more compact with slightly bigger grains. They are more densely packed and exhibit a lower concentration of pores. The sintering at 900°C (Fig. 3D) results in large LATP grains of ca. 4 µm in size. Moreover, in this case the ceramics' porosity is also considerably reduced.

Apparent density values (Table 1) determined by Archimedes method for LATP and LATP–LBSO vary in the narrow range of 2.66 – 2.77 g·cm$^{-1}$. The collected data suggests that the sintering of the LATP in the presence of LBSO densifies the material but the LBSO content in the composite has only marginal influence on the apparent density values. The most compact samples were obtained when heat-treated at 800°C and 900°C which is consistent with the reduced porosity observed in SEM images.

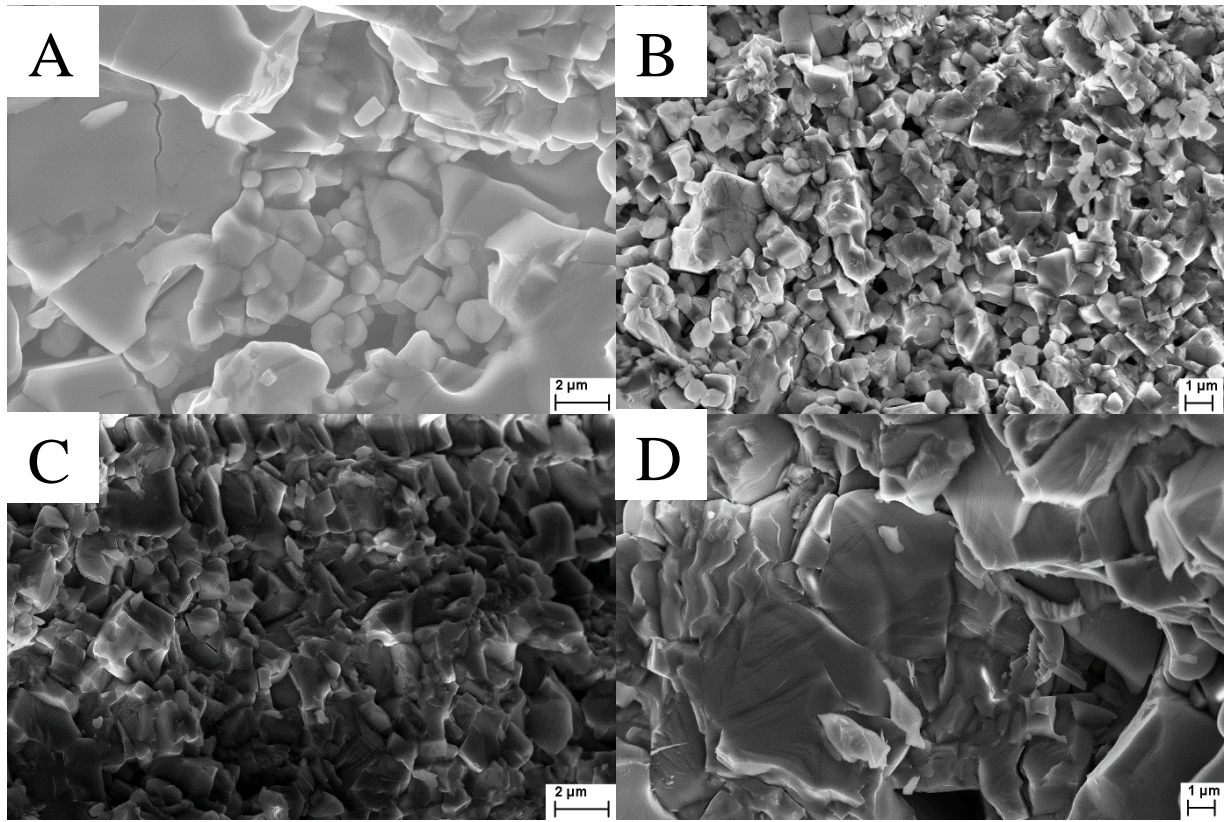

Figure 3 SEM images of the pellets of LATP sintered at 900°C (A), LATP–0.1LBSO sintered at 700°C (B), 800°C (C), 900°C (D).

Table 1 Values of the apparent densities of the materials sintered at different temperatures.

| composite | Sintering temperature [°C] | Apparent density [g·cm$^{-3}$] |
|---|---|---|
| LATP | 700 | 2.69 ± 0.03 |
| | 800 | 2.70 ± 0.03 |
| | 900 | 2.71 ± 0.04 |
| LATP–0.1LBSO | 700 | 2.66 ± 0.03 |
| | 800 | 2.77 ± 0.04 |
| | 900 | 2.76 ± 0.04 |
| LATP–0.2LBSO | 700 | 2.64 ± 0.04 |
| | 800 | 2.78 ± 0.04 |
| | 900 | 2.74 ± 0.03 |
| LATP–0.3LBSO | 700 | 2.66 ± 0.03 |
| | 800 | 2.76 ± 0.03 |
| | 900 | 2.74 ± 0.03 |

### 3.4 MAS NMR

MAS NMR studies were carried out on the LATP base material (non-sintered and sintered at 900°C) and LATP–0.3LBSO composite (non-sintered and sintered at 900°C).

### 3.4.1 $^{27}$Al MAS NMR

$^{27}$Al MAS NMR spectrum (Fig. 4) for the non-sintered LATP material consist of the three main peaks located at −15, 13 and 40 ppm. They were attributed to the hexa- ($AlO_6$), penta- ($AlO_5$) and tetra-coordinated ($AlO_4$) aluminum sites, respectively [8, 14, 19, 33, 35]. As $AlO_6$ sites are characteristic for aluminum coordination in the NASICON structure, then it is concluded that −15 ppm line could be associated with the LATP phase. $AlO_4$ signals are assigned to $AlPO_4$ material, which is also observed by XRD [14, 19, 33]. The $AlO_5$ resonance is produced by a glassy phase containing $Al^{3+}$ ions. After sintering at 900°C, the intensity of the $AlO_6$ resonance increased, while that of $AlO_5$ resonance is strongly reduced. The intensity of the peak at 40 ppm ascribed to $AlO_4$ sites in $AlPO_4$ remains the same. The deconvolution of the $^{27}$Al spectrum shows the presence of an additional $AlO_4$ peak at 31 ppm, which can be associated with some glassy phase containing aluminum ions. The integrated intensity of this peak is lower than that of $AlO_5$ signal in non-sintered LATP. That observation may indicate that during sintering of LATP, $Al^{3+}$ ions migrate from the glassy or disordered phases containing $AlO_5$ sites to the LATP structure and occupy their hexa-coordinated sites.

For the non-sintered LATP-LBSO composite, the relative integrated intensities of the lines at −15, 13 and 40 ppm do not significantly change when compared to the non-sintered LATP material. However, we observe an additional peak located at 57 ppm. It seems likely that this resonance is due to the $^{27}$Al nuclei in LBSO that migrated from an alumina crucible during the melting step of the glass preparation process. Conversely the sintering of composite at 900°C drastically modifies the $^{27}$Al spectrum. The lines ascribed to $AlO_4$ and $AlO_5$ disappear and only the AlO$_6$ signal of LATP remains. Hence, it may be concluded that during the sintering of LATP–LBSO composites, aluminum ions are transferred from AlPO$_4$, glassy phase containing Al$^{3+}$ ions and LBSO glass to the LATP. We assume that during this transformation, Al$^{3+}$ ions substitute Ti$^{4+}$ ions in the LATP structure. At the same time, Li$^+$ ions must also diffuse into LATP and occupy its cation vacancies so that the material remains electrically neutral. Furthermore, the presented MAS NMR data show that during heat-treatment of LATP–LBSO systems, no foreign (amorphous as well as crystalline) phases containing Al$^{3+}$ ions are formed.

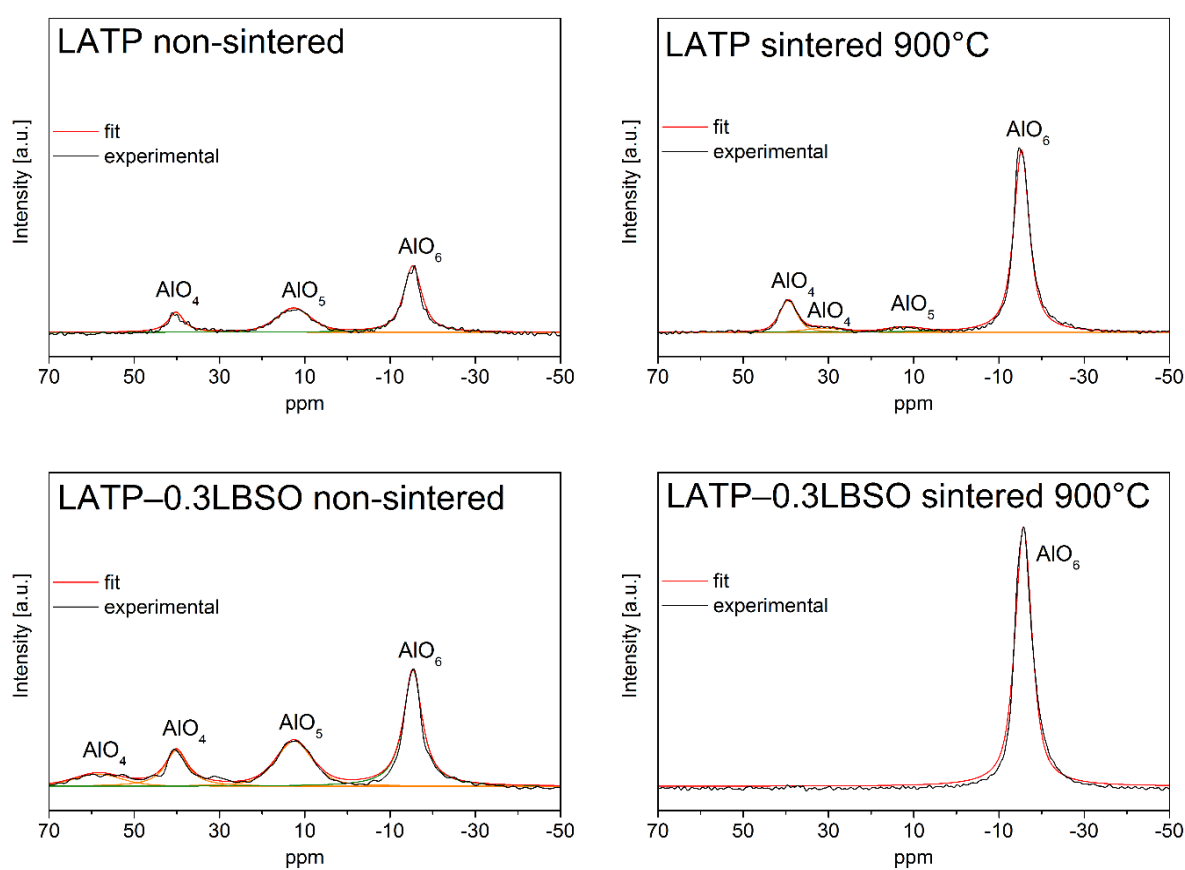

Figure 4 $^{27}$Al MAS NMR spectra of LATP non-sintered and sintered at 900°C as well as LATP–0.3LBSO non-sintered and sintered at 900°C. The experimental and simulated spectra are displayed as black and red lines, respectively. The simulated spectra are the sum of distinct lineshapes displayed as orange and green lines. The best fit parameters are given in Table S1. The spectra of the four samples are displayed with identical intensity scales.

### 3.4.2 $^{31}$P MAS NMR

Fig. 5 presents the $^{31}$P MAS NMR spectrum of LATP, which is dominated by an asymmetrically broadened signal with a maximum at −27.5 ppm. This signal can be simulated as the sum of three Gaussian lines with isotropic chemical shifts of −27.7, −27.0 and −26.3 ppm, which according to the literature, may correspond to different structural units of LATP phase: P(OTi)$_4$, P(OTi)$_3$(OAl)$_1$ and P(OTi)$_2$(OAl)$_2$, respectively [6, 8, 13]. Moreover, two weak lines at −24.4 and −18.8 ppm are observed. The first one was assigned to $^{31}$P nuclei in berlinite [36]. The resonance at −18.8 ppm can be assigned to $Q_3^0$ sites [P(OAl)$_3$O] present in the glassy phase containing AlO$_5$ sites [37].

After sintering, the integrated intensity of P(OTi)$_4$ decreases, whereas those of P(OTi)$_3$(OAl)$_1$ and P(OTi)$_2$(OAl)$_2$ remain the same and the signal of P(OTi)$_1$(OAl)$_3$ sites appears at −25.7 ppm. The intensity of the $^{31}$P berlinite signal also slightly increases. Furthermore, the resonance of $Q_3^0$ sites disappears, while we observe a new peak at −23.4 ppm. As this shift corresponds to [P(OTi)$_3$O] environment and this phase is not detected by XRD, it may be produced by a glassy titanophosphate phase [38]. These results are consistent with the $^{27}$Al data. During sintering, the glassy phase containing AlO$_5$ sites decomposes and the Al$^{3+}$ ions replace some of the Ti$^{4+}$ ions in LATP. The released Ti$^{4+}$ ions diffuse outside of LATP particles, where they form glassy titanophosphate phase with phosphate ions.

Fig. 5 also presents the deconvoluted spectra for the composite before and after sintering at 900°C. No significant difference between the spectra of non-sintered LATP and composite is observed. However, the $^{31}$P spectrum of sintered LATP–LBSO significantly differs from that of sintered LATP. First of all, three additional lines are observed at −9.8, −5.8 and 9.8 ppm and are ascribed to $^{31}$P nuclei in some crystalline phosphates, including LiTiPO$_5$, Li$_4$P$_2$O$_7$ (also observed by XRD) or/and Li$_3$PO$_4$ phases [39]. Secondly, as already noticed for sintered LATP, the integrated intensity of P(OTi)$_4$ signal decreases and the signal of P(OTi)$_1$(OAl)$_3$ sites

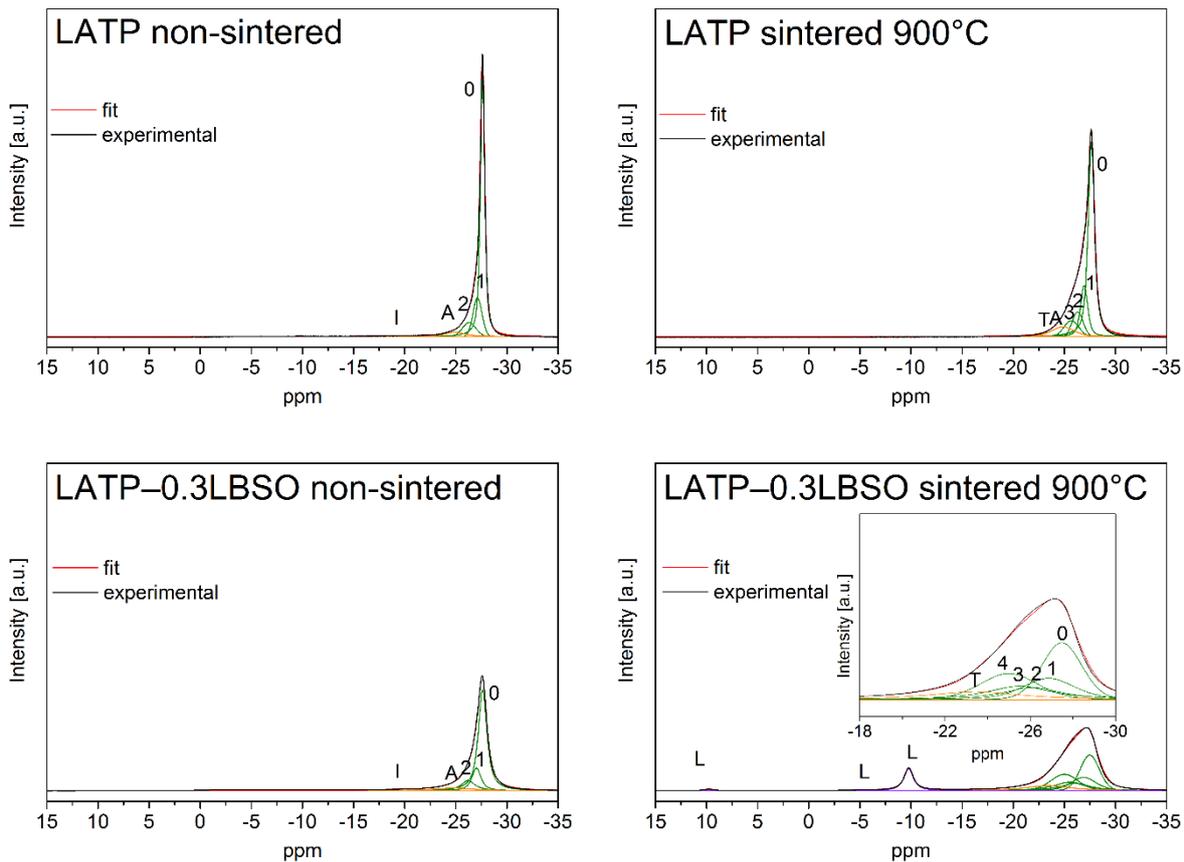

Figure 5 $^{31}$P MAS NMR spectra for LATP non-sintered and sintered at 900°C as well as LATP–0.3LBSO non-sintered and sintered at 900°C. The experimental and simulated spectra are displayed as black and red lines, respectively. The simulated spectra is the sum of distinct lineshapes displayed as orange, green and violet lines. Numbers x = 0-4 stands for P(OTi)$_{4-x}$(OAl)$_x$ sites, A for berlinite, T for amorphous titanophosphate phase, I for impurity phase and L for lithium conducting phosphates. The spectra of the four samples are displayed with identical intensity scales. The best fit parameters are given in Table S2.

appears. An additional peak at −25.1 ppm assigned to P(OAl)$_4$ is also detected. The presence of this peak and the larger relative integrated intensity of P(OTi)$_1$(OAl)$_3$ sites for sintered composite than for sintered LATP indicate a larger substitution of Ti$^{4+}$ ions by Al$^{3+}$ ones in the case of the sintered composite. This higher amount of defects in LATP explains the broadening of the P(OTi)$_{4-n}$(OAl)$_n$ resonances with $0 \leq n \leq 3$ in the sintered composite. Lastly, the peak at

−23.4 ppm assigned to glassy titanophosphate phase is more intense for the sintered composite than the sintered LATP. These results are consistent with the $^{27}$Al NMR data, which indicated the decomposition of the berlinite during the sintering of the composite. This decomposition results in the release of a larger amount of Al$^{3+}$ ions, a larger substitution of Ti$^{4+}$ ions by Al$^{3+}$ ones and the diffusion outside of LATP particles of a larger amount of Ti$^{4+}$ ions, which reacts at the grain surface with PO$_4^{3-}$ ions to form LiTiPO$_5$, Li$_4$P$_2$O$_7$ and Li$_3$PO$_4$ phases as well as amorphous titanophosphate phase.

### *3.4.3 $^7$Li MAS NMR*

The deconvoluted $^7$Li MAS NMR spectra for the as-prepared and sintered neat LATP is presented on Fig. 6. The former exhibits two overlapping lines at −1 and −0.8 ppm, with quadrupolar coupling constant ($C_Q$) equal to 38 and 71 kHz. They could be assigned to the Li1 and Li3 sites in NASICON crystal structure, respectively [8, 13, 35]. The relative integrated intensities of the lines indicates that the occupancies of Li1 and Li3 sites by Li atoms are approximately equal. After sintering, the $C_Q$ value of both $^7$Li sites decreases to 27 and 54 kHz for Li1 and Li3, respectively. The observed decrease may be related to the partial substitution of Ti$^{4+}$ ions by Al$^{3+}$ ions. The sintering also produces a broadening of the $^7$Li NMR signal, which results from the distribution of local environments produced by the substitution. In addition, the relative integrated intensity of Li3 site increases, which indicates that this site becomes the preferred one for Li$^+$ ions. For the composite, an additional peak at 1.8 ppm was observed and assigned to LBSO. Apart from that peak, the $^7$Li NMR spectra of LATP and composite are similar.

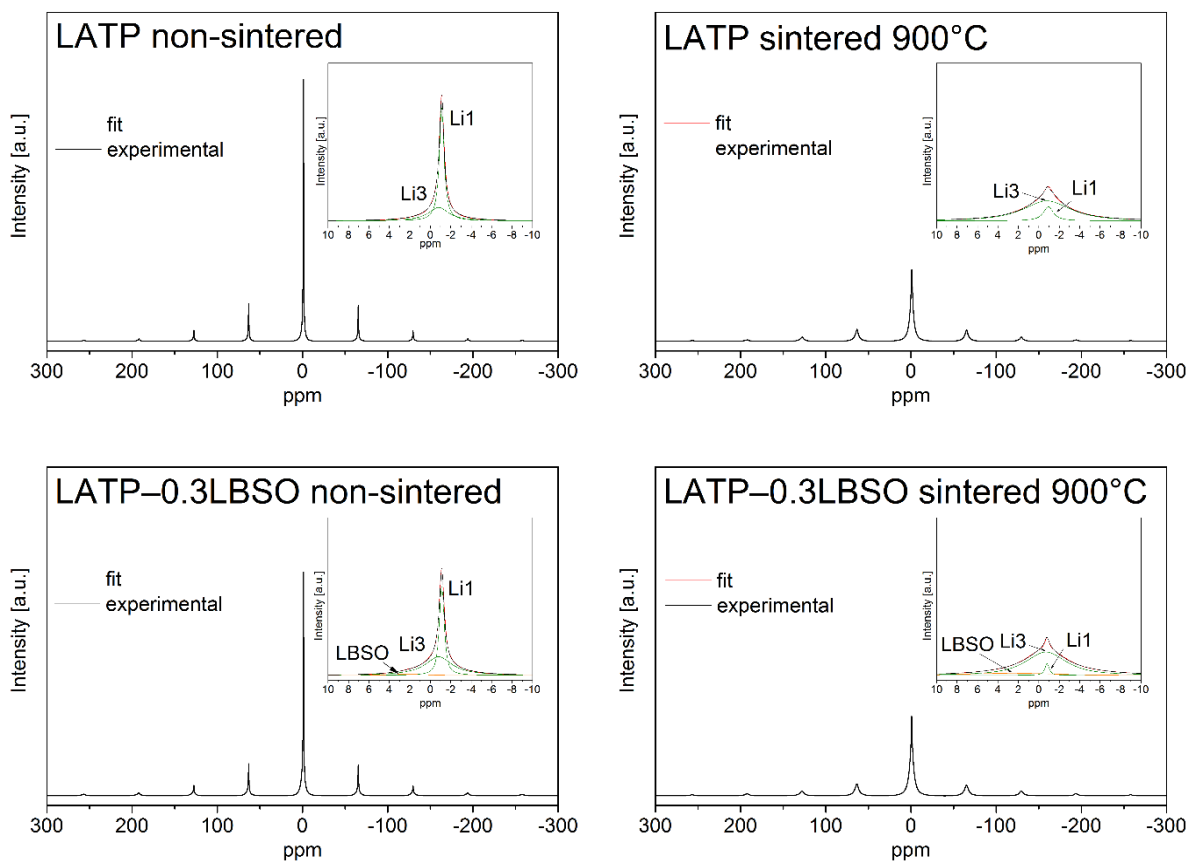

Figure 6 CT of $^7$Li NMR spectra under MAS conditions of LATP non-sintered and sintered at 900°C as well as LATP–0.3LBSO non-sintered and sintered at 900°C. The experimental and simulated spectra are displayed as black and red lines, respectively. The simulated spectra are the sum of distinct lineshapes displayed as orange and green lines. The different sites of LATP are labeled Li1 and Li3. The spectra of the four samples are displayed with identical intensity scales. The best fit parameters are given in Table S3.

### 3.4.4 $^{11}$B MAS NMR

Fig. 7 presents the $^{11}$B MAS NMR spectra for the composites before and after sintering. The spectrum for non-sintered composite consist of two Gaussian lines located at 16.1 and 0.8 ppm, which are ascribed to $BO_3$ and $BO_4$ sites, respectively. For the non-sintered material, the integrated intensity of the $BO_3$ site is larger than that of $BO_4$. Hence, the $BO_3$ units are more abundant than the $BO_4$ ones. From the composition of the LBSO glass and the molar fractions of $BO_3$ and $BO_4$ sites, we can determine by imposing the condition of charge neutrality that the $BO_3$ and $BO_4$ sites consist of $[BO_3]^{3-}$ and $[B(OB)_2O_2]^{3-}$ environments respectively [40]. $BO_3$ and $BO_4$ sites bonded to S atoms are not observed in the LBSO glass before sintering. After sintering, the integrated intensity of $BO_4$ signal become larger than that of $BO_3$. Hence, the $BO_4$ units become more abundant than the $BO_3$ ones. For LBSO glass, it has been shown that the amount of $BO_4$ units increases for increasing $x(S)$ and decreasing ratio between the mole fractions of Li and B atoms, $x(Li)/x(B)$, provided $x(Li)/x(B)$ remains greater than 0.5 [40]. For $Li_{2.9}B_{0.9}S_{0.1}O_{3.1}$, $x(Li)/x(B) = 3.2$ and the increased molar fraction of $BO_4$ is consistent with the diffusion of $Li^+$ ions from LBSO to LATP, which decreases $x(Li)/x(B)$ and increases $x(S)$. We can also notice a shift of both $BO_3$ and $BO_4$ signals toward negative values after sintering. These shift result from the formation of B−O−S bonds. Furthermore, both $BO_3$ and $BO_4$ signals can only be simulated using two lineshapes for each signal. The peak centered at 15.8 ppm is assigned to $[BO_3]^{3-}$ site, whereas the peak resonating at 10.5 ppm is produced by $[BO_2(OSO_3)]^{3-}$ sites produced by the formation of B−O−S bond between $SO_4^{2-}$ and $[BO_3]^{3-}$ anions. The peak centered at −0.9 ppm is assigned to $[B(OB)_2O_2]^{3-}$ site, whereas the peak at −3.3 ppm is ascribed to $[B(OB)_2O(OSO_3)]^{3-}$ sites produced by the formation of B−O−S bond between $SO_4^{2-}$ anions and $[B(OB)_2O_2]^{3-}$ site as well as $[BO_2(OSO_4)]^{5-}$ environments resulting from the coordination of S atom of $SO_4^{2-}$ anions by the non-bridging oxygen atom of $B(OB)_2O^-$ group.

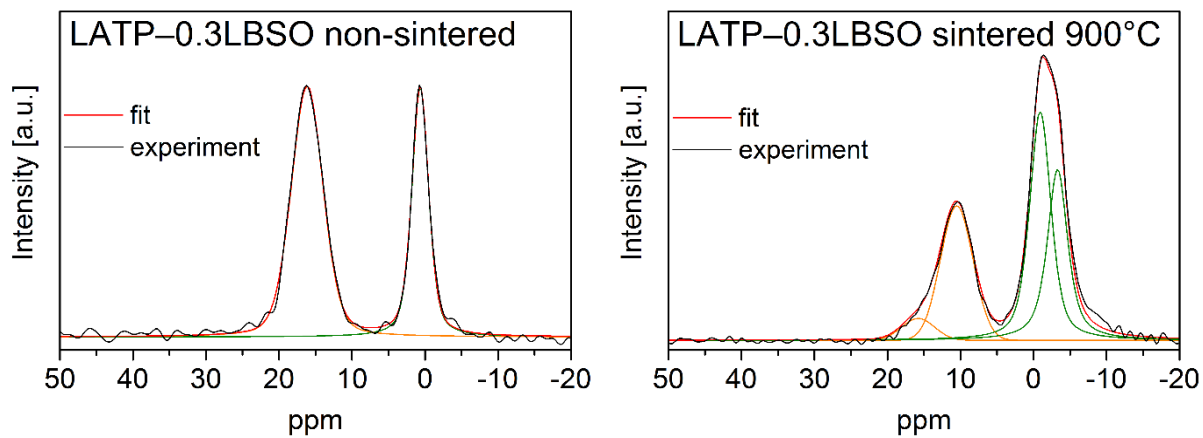

Figure 7 $^{11}$B MAS NMR spectra of LATP–0.3LBSO non-sintered and sintered at 900°C. The experimental and simulated spectra are displayed as black and red lines, respectively. The simulated spectra are the sum of distinct lineshapes displayed as green lines. The best fit parameters are given in Table S4.

*3.5 Impedance spectroscopy*

The electric properties of the materials under study were determined by means of impedance spectroscopy method and their results are presented in Figs. 8 and 9. The obtained impedance data collected at 30°C are shown on Nyquist plots (Fig. 8) for two typical exemplary cases: LATP sintered at 900°C (Fig. 8A) and LATP–0.1LBSO annealed at 700°C (Fig. 8B). The Nyquist plot for LATP base material consists of almost regular semicircle followed by a spur at lower frequencies. A more accurate analysis allows discerning another small semicircle at higher frequencies. Therefore, it may be concluded that the electrical properties of LATP could be modelled by an electrical circuit composed of two loops connected in series, each containing a resistor R shunted by a constant phase element CPE. This type of equivalent circuit is typical for materials, in which ionic transport takes place through two different phases: grain interior (at higher frequencies) and grain boundaries (at lower frequencies).

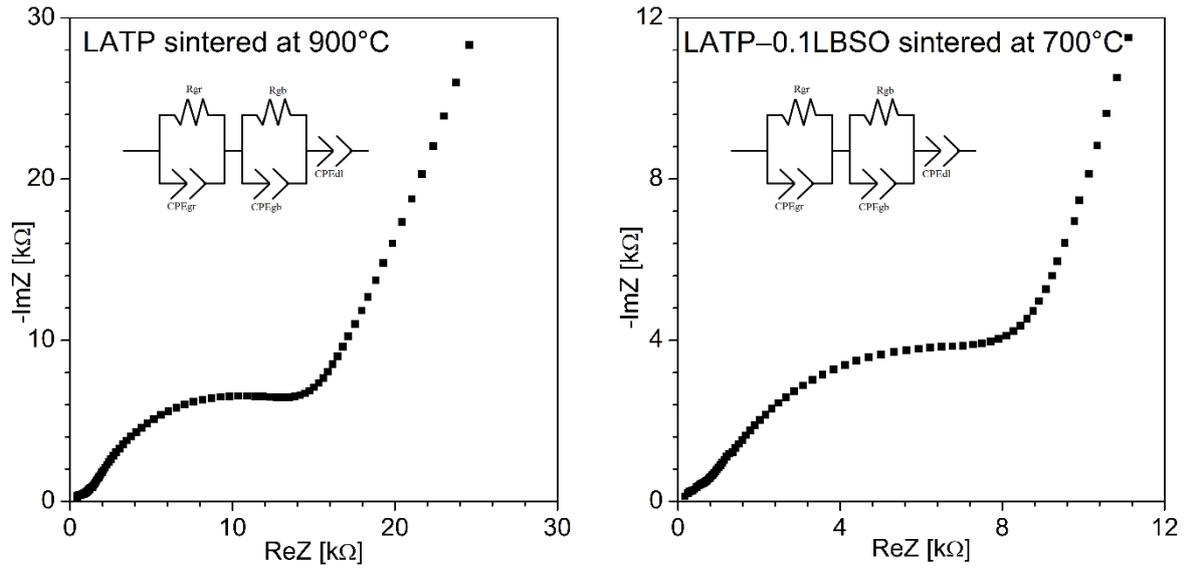

Figure 8 Nyquist plots for the data collected at 30°C for (A) LATP ceramic sintered at 900°C and (B) LATP–0.1LBSO composite sintered at 700°C. The equivalent circuits best modeling the electrical properties of the sintered materials are displayed as insets.

The apparent conductivity of each component could be derived using the formula $\sigma = L/(R \cdot A)$, where $L$ and $A$ stand for sample thickness and electrode area, respectively. The resistance $R$ is given by the intersection point of the arc with Re $Z$ axis. From the spectra, it is clearly visible that the resistance of the grain boundary region is much higher than that for the bulk. As a consequence its impact on the overall electrical properties of the material is predominant. Given the total resistance $R_{tot} = R_{gr} + R_{gb}$ and geometrical factor $L/A$, the total conductivity $\sigma_{tot}$ for ceramic LATP can be determined as ca. $4.65 \times 10^{-5}$ S·cm$^{-1}$.

The addition of LBSO glass into the LATP base matrix with subsequent sintering definitely enhanced its total ionic conductivity (Fig. 8B). The Nyquist plots for the sintered LATP–LBSO composites exhibit two semicircles followed by a spur. The values of the estimated apparent conductivity ($\sigma_{tot}$) are listed in Table 2. More detailed analysis of these results reveals some interesting correlations: (i) for the composites, $\sigma_{tot}$ slightly decreases with higher LBSO content, regardless of sintering temperature, (ii) sintering at 800°C yields composites with the highest grain as well as total conductivity values for each composition.

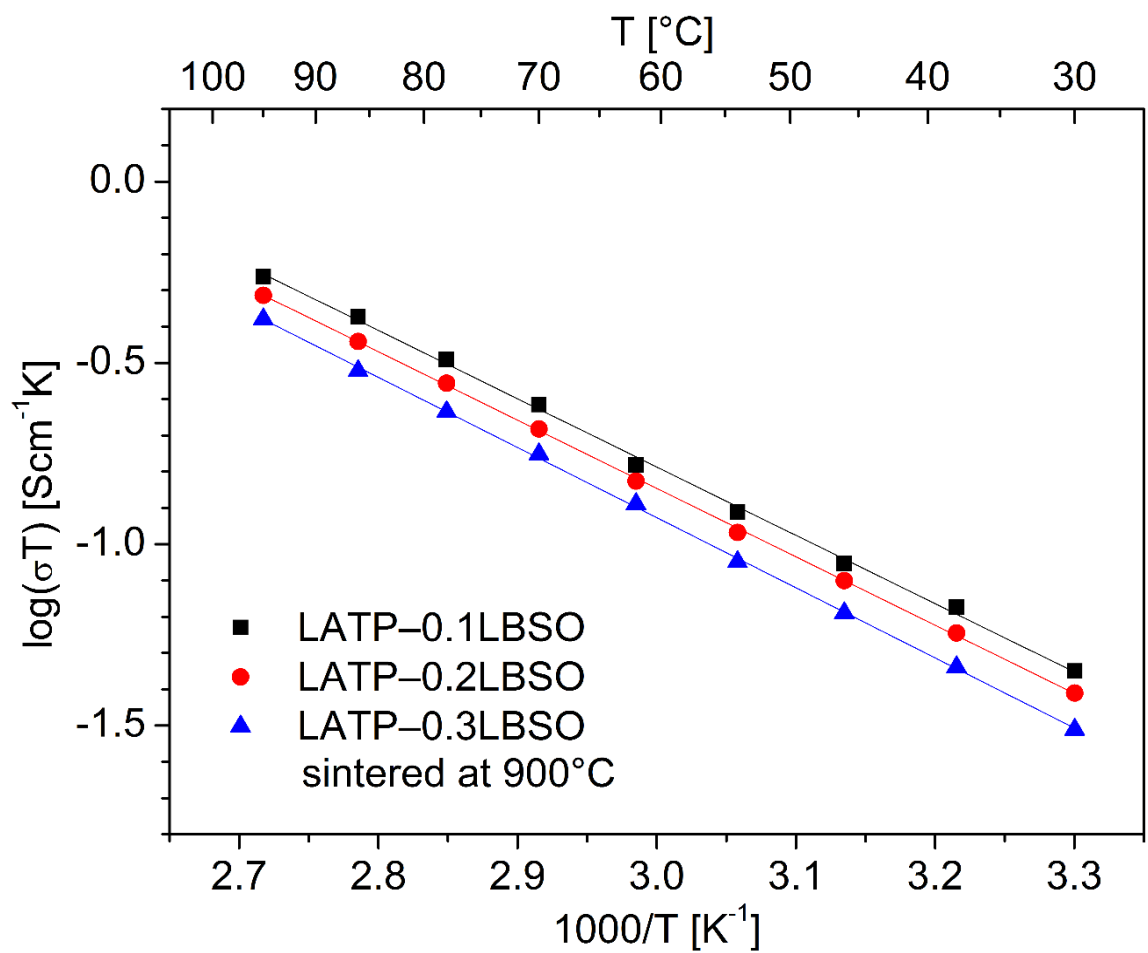

Figure 9 Arrhenius plots of the total electric conductivity of the LATP–LBSO composites sintered at 900°C.

The temperature dependent impedance spectroscopy results show for all composites, the total ionic conductivity fulfills Arrhenius dependence, regardless sintering temperature and composition (Fig. 9). Table 2 lists the values of bulk $E_{gr}$ and total $E_{tot}$ activation energy. It was noticed that for the LATP–LBSO composites, $E_{tot}$ and $E_{gr}$ vary in the narrow range from 0.38 to 0.41 eV and 0.29 to 0.31 eV, respectively. The lowest value of $E_{tot}$ was obtained for materials sintered at least at 800°C. It is worth mentioning that these results are consistent with the values reported in the literature, which vary in a comparable range for both $E_{gr}$ (0.29 [9], 0.28 [8], 0.22 eV [12]) and $E_{tot}$ (0.38 [9], 0.36 [8] and 0.3 eV [12]).

Table 2 Values of bulk, grain boundary and total ionic conductivities at 30°C as well as the bulk and total activation energies.

| composite | $T_{\text{sint}}$ [°C] | $\sigma_{\text{gr}}$ (30°C) [S·cm$^{-1}$] | $\sigma_{\text{gb}}$ (30°C) [S·cm$^{-1}$] | $\sigma_{\text{tot}}$ (30°C) [S·cm$^{-1}$] | $E_{\text{gr}}$ [eV] | $E_{\text{tot}}$ [eV] |
|---|---|---|---|---|---|---|
| LATP | 700 | 8.8 × 10$^{-5}$ | 2.0 × 10$^{-6}$ | 1.9 × 10$^{-6}$ | 0.21 | 0.46 |
|  | 800 | 2.0 × 10$^{-4}$ | 1.1 × 10$^{-5}$ | 1.0 × 10$^{-5}$ | 0.21 | 0.39 |
|  | 900 | 5.1 × 10$^{-4}$ | 5.2 × 10$^{-5}$ | 4.7 × 10$^{-5}$ | 0.21 | 0.4 |
| LATP–0.1LBSO | 700 | 5.6 × 10$^{-4}$ | 1.0 × 10$^{-4}$ | 8.8 × 10$^{-5}$ | 0.30 | 0.41 |
|  | 800 | 9.8 × 10$^{-4}$ | 1.8 × 10$^{-4}$ | 1.5 × 10$^{-4}$ | 0.31 | 0.39 |
|  | 900 | 7.2 × 10$^{-4}$ | 1.9 × 10$^{-4}$ | 1.5 × 10$^{-4}$ | 0.30 | 0.38 |
| LATP–0.2LBSO | 700 | 3.8 × 10$^{-4}$ | 5.6 × 10$^{-5}$ | 4.9 × 10$^{-5}$ | 0.29 | 0.41 |
|  | 800 | 8.1 × 10$^{-4}$ | 1.7 × 10$^{-4}$ | 1.4 × 10$^{-4}$ | 0.31 | 0.39 |
|  | 900 | 5.7 × 10$^{-4}$ | 1.7 × 10$^{-4}$ | 1.3 × 10$^{-4}$ | 0.29 | 0.38 |
| LATP–0.3LBSO | 700 | 3.0 × 10$^{-4}$ | 1.5 × 10$^{-5}$ | 1.4 × 10$^{-5}$ | 0.29 | 0.41 |
|  | 800 | 7.2 × 10$^{-4}$ | 1.5 × 10$^{-4}$ | 1.3 × 10$^{-4}$ | 0.31 | 0.39 |
|  | 900 | 5.5 × 10$^{-4}$ | 1.3 × 10$^{-4}$ | 1.0 × 10$^{-4}$ | 0.30 | 0.38 |

The ionic properties of LATP–LBSO composites may be analyzed in terms of the brick layer model (BLM) [41-46], where the ceramic material is represented as a system of adjacent cubes of size $D$, each coated with a thin layer of thickness $d$. Assuming that cubes correspond to LATP grains and the thin layer between them to the grain boundary region, the total conductivity, $\sigma_{tot}$ could be expressed as follows [41]:

$$\sigma_{\text{tot}} = \frac{(1-2\alpha)^2 \sigma_{\text{gr}}}{(1-2\alpha)+2\alpha\frac{\sigma_{\text{gr}}}{\sigma_{\text{gb}}}} + 4\alpha(1-\alpha)\sigma_{\text{gb}} \tag{1}$$

where, $\alpha = d/D$ and, $\sigma_{gr}$ and $\sigma_{gb}$ denote the true grain and grain boundary conductivity values but not the apparent ones, determined by means of impedance spectroscopy. If the resistivity of the grain boundary region is much higher than the resistivity of the bulk, i.e. $\sigma_{gr} >> \sigma_{gb}$, then the formula can be simplified into:

$$\sigma_{\text{tot}} \approx \left[\frac{(1-2\alpha)^2}{2\alpha} + 4\alpha(1-\alpha)\right]\sigma_{\text{gb}} \tag{2}$$

In that case, $\sigma_{tot} \approx \sigma_{gb}$ and the grain boundary conductivity becomes the limiting factor for the overall electric properties. Ceramic LATP as well as non–sintered composite meet that criterion, with $\sigma_{gr} > 10\sigma_{gb}$. Thus, the presence of some poorly conducting phases, such as berlinite and some glassy aluminophosphate phase, located between LATP grain reduces the conductivity of non-sintered LATP. As for sintered LATP, it exhibits better electrical properties, probably due to decomposition of disordered phase, while berlinite is still present. Finally, the addition of LBSO followed by sintering results in the decomposition of all highly resistant phases at grain boundaries, including both $AlPO_4$ and glassy aluminophosphate phase. These phases are replaced by lithium ion conducting phases, such as $LiTiPO_5$, $Li_4P_2O_7$ and $Li_3PO_4$. For that reason, the grain boundary conductivity significantly increases.

Considering the case, when the grain size is large or the layer is thin ($\alpha << 1$), Eq. (1) transforms into:

$$\sigma_{\text{tot}} = \frac{\sigma_{\text{gr}}}{1+2\alpha\frac{\sigma_{\text{gr}}}{\sigma_{\text{gb}}}} \tag{3}$$

If $\alpha\sigma_{gr}/\sigma_{gb} << 1$, i.e. grains are large and $\sigma_{gb}$ is comparable to $\sigma_{gr}$, the model predicts that $\sigma_{tot} \approx \sigma_{gr}$.

On the basis of the BLM and Eqs. 1-3, we suggest that at least two factors can provoke the enhancement of total ionic conductivity in LATP–LBSO composites. The first factor is related to the microstructure of the material, i.e. the size of the grains and the thickness of grain boundary layer. The second factor is the conductivity of the grain boundary phases. Both factors can affect the total conductivity simultaneously as the sintering process not only promotes the grain growth but also triggers changes in phase composition of grains and grain boundary region. Particularly, NMR studies show that the as-prepared LATP contains not only the NASICON-like phase, but also some amounts of by-products. Hence, the concentration of $Al^{3+}$ ions in NASICON phase should be lower than expected. The sintering of the as-prepared LATP allows the synthesis process to continue, yielding grains with the chemical formula closer to the assumed one. During sintering, many simultaneous processes also occur, including the decomposition of inter-grain phases with subsequent release of aluminium ions, the substitution of $Ti^{4+}$ by $Al^{3+}$ ions in the NASICON crystal lattice and the reaction of the released $Ti^{4+}$ atoms with the decomposition by-products resulting in formation of new phases at grain boundaries.

The impact of other phases on total ionic conductivity could be clearly visible considering different sintering temperatures and compositions. It is worth reminding that composites under study exhibit nearly the same values of grain and total activation energy. The values of total ionic conductivity for a given composition is the lowest at 700°C and decreases with increasing LBSO content, whereas for a given composition, the changes in total ionic conductivity are barely noticeable after sintering at 800 or 900°C. Considering these results in terms of BLM model and SEM investigations, it may be concluded that the value of total ionic conductivity (for a given composition) increases most probably due to the grain growth, indicating that $α(800$ or $900°C) < α(700°C)$, but the variation of conductivity with LBSO content strongly suggests that limiting the amount of foreign phases in the composite is essential to obtain a good ceramic lithium ion conductor.

## 4. Conclusions

The study on the impact of glassy additive on the structural and electrical properties of the $Li_{1.3}Al_{0.3}Ti_{1.7}(PO_4)_3$–$xLi_{2.9}B_{0.9}S_{0.1}O_{3.1}$ ($0 \leq x \leq 0.3$) system, clearly shows that the formation of composite followed by the appropriate heat treatment is an efficient way to obtain a highly conducting solid electrolyte for lithium ion batteries. For the system under study, three processes are identified as probably responsible for the enhancement of total ionic conductivity: (i) the growth of grains, (ii) the decomposition of berlinite and amorphous aluminophosphate phases at the grain boundaries and (iii) the formation of some lithium conducting phosphates in the presence of $Li_{2.9}B_{0.9}S_{0.1}O_{3.1}$ glass. Among composites under study, the best conducting one, LATP–0.1LBSO sintered at 800°C, exhibits $\sigma_{tot} = 1.5 \times 10^{-4}$ S·cm$^{-1}$ at room temperature, almost one order of magnitude higher than ceramic LATP. The ease of preparation as well as good electrical properties of LATP-LBSO composites are the main factors advocating for their application as solid electrolytes in the all-solid lithium ion battery technology. Their performance in prototype electrochemical cells will be the subject of our further study.

**Acknowledgments**

This project has received funding from the European Union's Horizon 2020 research and innovation program under grant agreement No 731019 (EUSMI).